\documentclass[aps, floats, epsf,twocolumn]{revtex4}
\usepackage{graphics}

\def\d{{\mathrm{d}}}

\begin{document}

\title{Permanent confinement in compact QED$_3$ with fermionic matter}

\author{Igor F. Herbut and Babak H. Seradjeh}

\address{Department of Physics, Simon Fraser University,
Burnaby, British Columbia, Canada V5A 1S6 }

\begin{abstract}
We argue that the compact three dimensional electrodynamics
with massless relativistic fermions is always in the confined phase,
in spite of the bare interaction between the magnetic
monopoles being rendered logarithmic by fermions. The effect is
caused by screening by other dipoles, which transforms the logarithmic
back into the Coulomb interaction at large distances. Possible
implications for the chiral symmetry breaking for fermions are discussed,
and the global phase diagram of the theory is proposed.
\end{abstract}

\maketitle

It is well known that the pure compact electrodynamics in three dimensions (3D) is
always in the confined phase, i. e. in the phase where magnetic monopoles form a
neutral plasma of free magnetic charges~\cite{polyakov}. When the gauge field is
coupled to matter, however, new possibilities emerge. In particular,
Abelian gauge fields coupled to massless relativistic
fermions (cQED$_3$) arise naturally in several theories of
high temperature superconductivity~\cite{kim},~\cite{herbut1}. cQED$_3$
also represents maybe the simplest theory that should contain the physics
of confinement and chiral symmetry breaking, and as such it may
be used as a valuable toy model for the QCD. It has been argued that
coupling to massless relativistic
fermions turns the interaction between monopoles from $1/x$ to $-\ln(x)$
at large distances $x$~\cite{ioffe},~\cite{marston},
~\cite{wen}, ~\cite{kleinert}, so that the
deconfined phase with bound monopole-antimonopole pairs may become stable at low
(effective) temperatures, in close analogy to the Kosterlitz-Thouless (KT)
transition. If so, this would suggest that, at least for some values of
coupling constants, compactness of the gauge field may be neglected at large
distances, and that the results pertaining to the continuum QED$_3$~\cite{maris} are
likely to remain valid. In particular, the chiral symmetry breaking instability of the
continuum QED$_3$ that has been linked to antiferromagnetism~\cite{herbut3} in
cuprates might then be expected to survive essentially intact.

\begin{figure}[t]
{\centering\resizebox*{80mm}{!}{\includegraphics{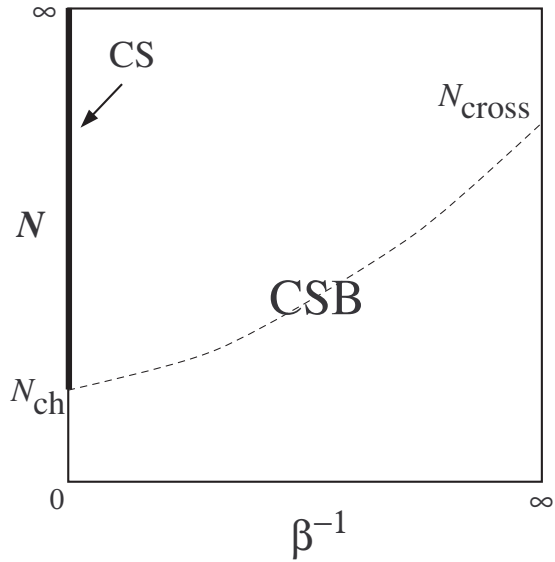}} \par}
  \caption[]{The proposed phase diagram of the cQED$_{3}$. True phase transition
  between the chirally symmetric (CS) and chiral symmetry broken (CSB) phases of
  fermions occurs only in the continuum limit ($\beta\to\infty$).
  Monopoles are in the plasma phase at all $\beta<\infty$ and $N<\infty$.
  Dashed line marks a crossover that corresponds to the
  $N_{\mathrm{cross} }(\beta)$, as discussed in the text.}
  \label{phase}
\end{figure}

In this paper we consider the effects of finite density of monopole-antimonopole
pairs (dipoles) on the putative, KT-like, confinement-deconfinement
transition in the cQED$_3$. We find that, in
stark contrast to two dimensions (2D), screening by logarithmically interacting
magnetic charges bound to dipoles in 3D always alters the {\it form} of the
interaction between the charges, from $-\ln(x)$ at large distance, to $1/x$. We
demonstrate this first by an elementary electrostatic argument, and then support it by
a controlled momentum shell renormalization group on the equivalent sine-Gordon
theory. As a result, largest dipoles are always unbound, and would be
expected to provide linear confinement of
electric charges. Finally, we argue that the absence
of the phase transition in the cQED$_3$ in terms of magnetic monopoles should imply the
same in terms of fermions. Relying on the existing numerical results and
qualitative arguments, we suggest that
the phase of free monopoles corresponds to the (confined) phase of
broken chiral symmetry for fermions, and that the (deconfined) 
chirally symmetric phase exists only in the continuum limit, and 
above a certain number of fermion species~\cite{maris}. The conjectured phase diagram
is presented in Fig.~\ref{phase}.

  We are interested in the cQED$_3$ defined on a cubic lattice:
  \begin{equation}
  S= S_F (\chi, \theta_\mu) - \beta \sum_{x,\mu,\nu} \cos( F_{\mu \nu} ),
  \label{lcQED}
  \end{equation}
where $F_{\mu \nu}= \epsilon_{\mu \nu\beta}\epsilon_{\beta\rho\sigma}
\Delta_\rho \theta _\sigma$ is the lattice version of the electromagnetic
tensor, and the coupling to fermions on a lattice may be written as
  \begin{eqnarray}
  S_F = \frac{1}{2} \sum_{x, \nu} \sum_{i=1}^{N/2}
  \eta_\nu (x) [ \bar{\chi}_i (x) e^{i \theta _\nu (x)}
  \chi_i (x+\hat{\nu}) \\ \nonumber
  - \bar{\chi}_i (x+\hat{\nu}) e^{-i \theta _\nu (x)} \chi_i (x) ],
  \end{eqnarray}
using the standard staggered fermions. Here, $x$ denotes the lattice sites
and $\hat{\nu}$ the direction of the links. In the continuum, $S_F$
corresponds to $N$ species of four component Dirac fermions~\cite{burkitt}.

The cQED$_3$ has been previously studied numerically in ~\cite{fiebig}.
Here, to make progress analytically and
following~\cite{kleinert}, we approximate~(\ref{lcQED}) with
 \begin{eqnarray}
 S \approx \frac{1}{2} \sum_{x,\nu,\mu}[F_{\mu \nu}(x) - 2\pi n_{\mu\nu}(x)]
           (\beta + \frac{N}{16 |\Delta|} ) \label{applcQED}\\\nonumber
           [F_{\mu\nu}(x) - 2\pi n_{\mu\nu}(x)]&&,
 \end{eqnarray}
where $1/|\Delta|$ should be understood as an inverse of the square root of the
lattice gradient squared. The term proportional to the coupling $\beta$ in Eq.~(\ref{applcQED})
is the standard Villain approximation to the second term in~(\ref{lcQED}). The form of
the term proportional to $N$ in~(\ref{applcQED}) is motivated by the
known continuum limit of the fermion polarization: neglecting compactness
of $a_\mu$ by setting
$n_{\mu\nu}(x)\equiv 0$ would give the standard one-loop Gaussian action
in the continuum,
with the correct coefficient~\cite{maris}. Just like in the $\sim \beta$
term, here we also retained only the leading power of fields and their
lattice derivatives. Eq. (3) may be understood as the quadratic  
approximation to cQED$_3$ in Eq. (1).
It leads to the expected logarithmic interaction
between the magnetic monopoles, and therefore appears to contain
the essential physics of cQED$_3$ we wish to address.

Standard duality transformations~\cite{einhorn} on~(\ref{applcQED}) lead to the
sine-Gordon action in the continuum
  \begin{equation}
  S_{\mathrm{sG}}= \int \d^3 \vec{x}
  \left[ \frac{(\vec{\nabla} \Phi(\vec{x})) |\vec{\nabla}| (\vec{\nabla} \Phi(\vec{x}))}
  {\pi^2(N + 16 \beta|\vec{\nabla}|) } - 2y \cos \Phi(\vec{x}) \right],
  \label{sG}
  \end{equation}
where $y \ll 1 $ is the fugacity of monopoles, and $\vec{x}$ is now the position vector in
the continuum. In presence of fermions ($N\neq 0$), at small momenta one can
expand in the original gauge coupling $\beta$ in Eq. (4),
which becomes a coefficient of the term quartic in derivatives, and thus
irrelevant by power counting.
We may therefore safely set it to zero, and
return to its (non-universal) effects on the phase diagram later.
We also find it useful to have an alternative representation of the
partition function defined by $S_{\mathrm{sG}}$,
  \begin{equation}
  Z= \sum_{n=0}^{\infty} \frac{y^n}{n!} \int \prod_{i=1}^n \d^3 \vec{x}_i \exp\left[
  -\frac{1}{2T } \sum_{i,j} q_i q_j V(|\vec{x}_i - \vec{x}_j|) \right] ,
  \end{equation}
in terms of unit magnetic charges $q_i = \pm 1$ 
at an effective temperature $T= 4/N
+O(\beta)$, interacting at large  distance via $V(x) = -\ln(x\Lambda)$. $\Lambda$ is
the ultraviolet cutoff implicit in Eq. (4). The sine-Gordon field
in Eq. (4) is then
$\langle \Phi(\vec{x}) \rangle = (i/T) \sum_i q_i \langle
V(|\vec{x}-\vec{x}_i|) \rangle $, where 
$\langle ... \rangle$ denotes averaging with respect to $Z$.  

 The excess of free energy due to a {\it single} isolated monopole in the sample
of a linear size $L$ is therefore
  \begin{equation}
  \Delta F = \frac{1}{2} \ln L - T \ln L^3,
  \label{f}
  \end{equation}
so that for $T < 1/6$ monopoles would be expected to be bound in pairs. This
corresponds to a ``critical'' number of fermion components $N =
24$~\cite{ioffe},~\cite{wen},
~\cite{kleinert}.  This argument neglects the effect of other
dipoles, however. In the standard KT transition their presence does not change the
critical temperature obtained by the simple energy-entropy argument, since the KT
fixed point lies at zero fugacity~\cite{kosterlitz}. We show next that the
situation in 3D is radically different.

Let us assume a distribution of charges  $\rho(\vec{x})$
interacting via $-\ln(x)$
interactions in 3D, located in a region of a finite size $R$. At a distance $x \gg R$
we can write the interaction as
  \begin{equation}
  V(\vec{x}) = \int \d^3 \vec{x}'\rho(\vec{x}') \left(-\ln|\vec{x}| +
   \frac{\vec{x} \cdot \vec{x}'}{x^2} + ...\right)
  \end{equation}
in spirit of multipole expansion~\cite{jackson}. Thus, for a medium with a
monopole
density $\rho(x)$ and a dipole moment density $\vec{P}(\vec{x})=\rho(\vec{x})\vec{x}$ , the
potential is given by
  \begin{equation}
  V(\vec{x}) = \int \d^3 \vec{x}' (-\ln|\vec{x}-\vec{x}'|)
  [ \rho(\vec{x}') - \vec{\nabla}' \cdot \vec{P} (\vec{x}')+...].
 \end{equation}
The energy of a small dipole with a moment $\vec{P} = q \vec{r}$ ($q=1$)
at $\vec{x}$ in a weak potential $V(\vec{x})$ is still
$\vec{P} \cdot \nabla V(\vec{x})$. Assuming non-interacting
dilute dipoles at a temperature $T$ then gives the average dipole moment
at $\vec{x}$
\begin{equation}
\langle \vec{P}(\vec{x}) \rangle =
\frac{ \langle r^2 \rangle}{3T} (- \nabla V(\vec{x}) ).
\end{equation}
Since the density of dipoles is proportional to $y^2$,
the polarizability of the medium is thus   
  \begin{equation}
  \chi \sim \frac{ \langle r^2 \rangle }{3T} y^2,
  \end{equation}
where the constant of proportionality depends on the
precise form of the short distance regularization of the
logarithmic interaction.

Writing Eq. (8) in Fourier space and combining Eqs. (9) and (10) into 
$\vec{P}(\vec{q}) = - i \chi \vec{q} V(\vec{q})$, we finally obtain 
the interaction due to an external point charge $Q$ in presence of
a finite density of dipoles 
  \begin{equation}
  V(\vec{q}) = \frac{Q}{|\vec{q}|^3 + \chi q^2},
  \end{equation}
or, in real space
  \begin{equation}
  V(\vec{x})= \frac{1}{4 \pi \chi} \frac{1}{x} + O(\frac{1}{x^2}).
  \end{equation}
This implies that in presence of other dipoles the first (energy) term in
Eq.~(\ref{f}) scales like $1/L$, so that the entropy always
dominates, and the largest dipoles should dissociate into free monopoles.
The `dielectric' phase of
logarithmically interacting monopoles in 3D thus, unlike in 2D,
appears destabilized by
dipole screening, as anticipated in ~\cite{subir}.
A similar conclusion has apparently also been drawn before in
the study of instantons in quantum antiferromagnets~\cite{murthy}.

\begin{figure}[t]
{\centering\resizebox*{80mm}{!}{\includegraphics{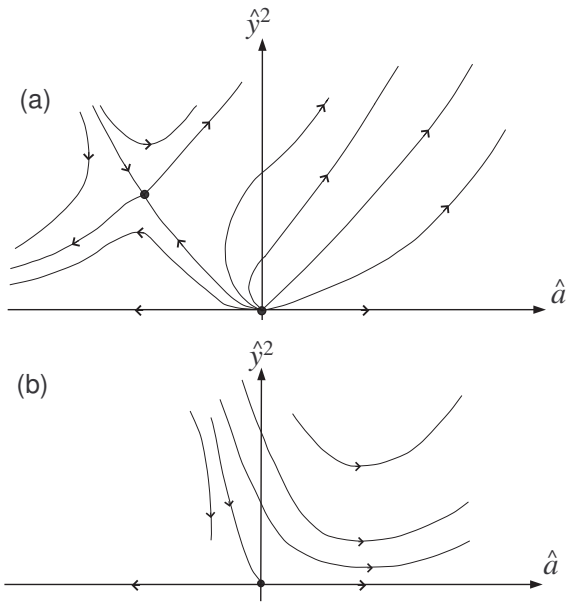}} \par}
  \caption[]{The flow diagram in the $\hat{y}^2$-$\hat{a}$ plane
  for $N < N_{cross} $ (a), and $N>N_{cross}$ (b). The non-trivial
  fixed point in (a) moved to lower right quadrant (not shown) in (b).}
  \label{flow01}
\end{figure}

The renormalization group treatment of the equivalent sine-Gordon theory closely
parallels the electrostatic consideration. Integrating out the modes of
$\Phi(\vec{x})$ in
Eq.~(\ref{sG}) with momenta $\Lambda/b < q < \Lambda$ one finds that the original
couplings are renormalized as
  \begin{eqnarray}
  N(b) &=& N+O(y^3),\\
  y(b) &=& b^3 y e^{-\frac{1}{2} G_> (0)}+O(y^3),\label{y}\\
  a(b) &=& b \left[a+ \frac{1}{3} y^2  e^{-G_> (0)} \int \d^3x\;
           x^2 (e^{G_>(x)}-1)\right]\nonumber\\
       & & +O(y^3), \label{a}
\end{eqnarray}
where $a$ is the coefficient in front of the
$(\vec{\nabla} \Phi(\vec{x} ))^2$ term in
$S_{\mathrm{sG}}$, Eq.~(\ref{sG}). Such a term is initially absent,
but gets {\it generated} to the second order in $y$.
Formally this is just what one expects, since
there is no particular symmetry in the Eq.~(\ref{sG}) that would prohibit its
appearance. Physically, as we have seen, this simply expresses the screening effect of
smaller dipoles onto the interaction between the monopoles in a larger dipole.
This term was effectively overlooked in the previous work on cQED$_3$
\cite{ioffe}, \cite{marston}, \cite{wen}, \cite{kleinert}, where
the authors consequently arrived at a different final result.

The correlation function appearing in the recursion relations~(\ref{y}) and~(\ref{a})
is
  \begin{equation}
  G_> (\vec{x}) = \int_{\Lambda/b<|\vec{q}|<\Lambda} \frac{\d^3 \vec{q}}{(2\pi)^3}
  \frac{\pi^2N  e^{i\vec{q} \cdot \vec{x}}}{2|\vec{q}|^3  + \pi^2N a q^2},
  \end{equation}
and requires some care in evaluation. Assuming  a low cutoff, or the effective
temperatures $T\Lambda^3 \ll y $, we may rewrite it as
  \begin{eqnarray}
  G_> (\vec{x}) = \frac{\pi^2N}{2\Lambda + \pi^2N a}  \\ \nonumber 
  \int \frac{\d^3 \vec{q}}{(2\pi)^3}
  \frac{e^{i\vec{q} \cdot \vec{x}} } {q^2}
  ( \frac{q^2}{q^2 + (\Lambda/b)^2}- \frac{q^2}{q^2 + \Lambda ^2}) ,
  \end{eqnarray}
where the integration is now unconstrained, but we have introduced a smoothing
function $q^2/(q^2 + \Lambda^2)$ instead of the sharp cutoff~\cite{jasnow}. We have
checked that when applied to the standard sine-Gordon problem in 2D, our
regularization scheme gives the correct anomalous dimension at the KT
transition. With this regularization we find
  \begin{equation}
  G_> (\vec{x}) = \frac1{4\pi}\frac{\pi^2N \ln(b)}{2+\pi^2N \hat{a}}e^{-x\Lambda} + O((\ln b)^2),
  \end{equation}
and therefore
  \begin{eqnarray}
  \dot{N}       &=& 0 + O(\hat{y}^3),\label{Ndot}\\
  \dot{\hat{y}} &=& \left[3- \frac{\pi N }{8 (2+\pi^2 N \hat{a})}\right] \hat{y} + O(\hat{y}^3),\\
  \dot{\hat{a}} &=& \hat{a}+ \frac{8\pi^2 N}{2+\pi^2 N \hat{a}}
  \hat{y}^2 + O(\hat{y}^3),\label{adot}
  \end{eqnarray}
where $\hat{y}=y / \Lambda^3$ and $\hat{a}=a/\Lambda$,
are the dimensionless couplings,
and $\dot{z}\equiv \d z/\d\ln b|_{b=1}$.

Two important features of the flow equations~(\ref{Ndot})-(\ref{adot}) should be
noted: 1) the relevant coupling $a$, even if absent initially, becomes generated at lower
cutoffs, and 2) $N$ is marginal to $O(y^3)$. We suspect $N$ to be an exactly marginal
coupling, since the action for slow modes should be analytic in $(q/\Lambda)^2$, so
the coefficient of the non-analytic $|q|^3$ term can not get renormalized to any
order~\cite{herbut2}.

The flow in the $\hat{y}^2$-$\hat{a}$ plane for $N < N_{\mathrm{cross}}$ and $N >
N_{\mathrm{cross}}$, $N_{\mathrm{cross}} = 48/\pi$, is depicted in
Fig.~\ref{flow01}. Starting from $\hat{a}=0, \hat{y}=\hat{y}_0$, the
fugacity in the former case monotonically increases as $b\rightarrow \infty$, while in
the latter it begins to increase only for $b>b^*$, with $b^*(\hat{y}_0) \approx
1/(96\pi \hat{y}_0) $ at $N \gg 1$. Nevertheless, since $\hat{y}=0$ is an
invariant line under the scale transformations,
fugacity must increase at large enough $b$ for any initial value. We
interpret this upward flow of the fugacity as an indication that monopoles are always
in the plasma phase.

Interestingly, massless fermions still make themselves felt in such a plasma,
since the non-analytic $|\vec{\nabla}|^3$ term in Eq. (4) translates into
a power-law behavior of the screened interaction at long distances.
Instead of the expected Debye-H\"uckel exponential decay, in the quadratic
approximation to (4) we find
$V_{\mathrm{scr} }(x) \approx - 6/(\pi ^4 y^2 N |\vec{x}| ^6)$ at large $x$.
The reader may recall a similar phenomenon
of Friedel oscillations in metals, where a non-analyticity due to
the Fermi surface produces a power-law behavior of the screened interaction
as well. In the present case the sign of the above power-law term
indicates {\it over-screening}, presumably
due to the extremely long range of the bare (logarithmic) interaction
between monopoles.

Let us next turn to possible implications of our findings for fermions. First, the
dimensionless coupling $\beta=1/(e^2 l)$, where $e$ is the dimensionfull charge, and
$l$ the lattice spacing. The lattice gauge field $\theta_\mu (x) = e l A_\mu (x)$,
where $A_\mu (x)$ is the gauge field in the continuum. The continuum limit is achieved
by keeping $e$ constant and taking $l \rightarrow 0$. In other words, $\beta
\rightarrow \infty$ should naively correspond to the continuum QED$_3$. In this theory chiral
symmetry for fermions is expected to be broken, and the condensate $\langle
\bar{\Psi} \Psi \rangle\neq 0$,  for $N<N_{\mathrm{ch}}$, where $N_{\mathrm{ch}}$ is of
order unity~\cite{maris},~\cite{appelquist},~\cite{hands}. On the other hand, for
$\beta < \infty $ we argued that once fermions are integrated out, the theory has no
phase transition. If correct, this implies that, had we proceeded in reverse and
integrated out everything but the fermions, we would have had to
find them in a single phase at any finite $\beta$. Relying on numerical
evidence that free monopoles seem to
{\it enhance} chiral symmetry breaking~\cite{fiebig}, we propose that it is
the phase with broken chiral symmetry that survives a finite $\beta$.

It was found previously~\cite{kocic} that the critical $N$ for chiral symmetry
breaking at $\beta=0$ is significantly larger than $N_{\mathrm{ch}}$ in the
continuum limit. We expect that for $\beta=0$, there 
is no true phase transition in the cQED$_3$, just the crossover occurring at
$N_{\mathrm{cross} }$. Making the simplest assumption that $\beta$
always stays irrelevant,
the same would hold at any $\beta <\infty$, but with the crossover line
shifted to $N_{\mathrm{cross}}(\beta)= N_{\mathrm{cross}}
- 16 \beta \Lambda + O((\beta \Lambda)^2)$.
The conjectured phase diagram that summarizes this discussion is presented in
Fig.~\ref{phase}. $\langle \bar{\Psi} \Psi \rangle\neq 0$ everywhere,
except on the line $1/\beta =0$, $N>N_{\mathrm{ch}}$.

To conclude, we argued that magnetic monopoles are always free in the
cQED$_3$ with gapless relativistic fermions, and the theory is thus
expected to be permanently confining. We proposed the phase diagram,
and conjectured that chiral symmetry for fermions is always spontaneously
broken. Our results seem to be in qualitative agreement with the early numerical results
of ref. \cite{luo}. They can also be generalized to the
case with non-relativistic fermions with Fermi surface \cite{igor-subir}. 
Although we have studied only the U(1) theory coupled to fermions in this
paper, similar arguments should be applicable to the case of 
critical scalar field \cite{kleinert}, \cite{chernodub}. 

We thank F. Nogueira and A. Sudbo for useful discussions, and S. Sachdev
for remarks that in part stimulated this work. This
work was supported by NSERC of Canada and the Research Corporation.

\end{document}